\documentclass[12pt]{article}

\usepackage{epsf,a4wide}

\renewcommand{\in}{\raise -3pt\hbox{\scriptsize in}}
\newcommand{\out}{\raise -3pt\hbox{\scriptsize out}}

\newcommand{\PP}{P}        % in case one needs to change notation
\newcommand{\mm}{m}
\newcommand{\dist}{\Phi}

\newcommand{\eqcm}{\: ,}   % punctuation in equations
\newcommand{\eqpt}{\: .}

\begin{document}

\begin{flushright}
DAPNIA--SPHN--98--34 \\
CPHT--S612--0598
\end{flushright}

\begin{center}
\vskip 3.5\baselineskip
\textbf{\Large Probing partonic structure in $\gamma^\ast \gamma
\to\pi \pi$ near threshold}
\vskip 2.5\baselineskip
M. Diehl$^{1}$, T. Gousset$^{2}$, B. Pire$^{3}$ and O. Teryaev$^{3,4}$
\vskip \baselineskip
1. DAPNIA/SPhN, CEA/Saclay, 91191 Gif sur Yvette, France \\
2. SUBATECH\,\footnote{Unit\'e mixte 6457 de l'Universit\'e de
   Nantes, de l'Ecole des Mines de Nantes et de l'IN2P3/CNRS},
   B.P. 20722, 44307 Nantes, France \\ 
3. CPhT\,\footnote{Unit\'e mixte C7066 du CNRS}, Ecole
   Polytechnique, 91128 Palaiseau, France \\ 
4. Bogoliubov Laboratory of Theoretical Physics, JINR, \\
   141980 Dubna, Moscow region, Russia\,\footnote{Permanent address}
\vskip 3\baselineskip
\textbf{Abstract} \\[0.5\baselineskip]
\parbox{0.9\textwidth}{
  Hadron pair production $\gamma^\ast \gamma \to h \bar h$ in the
  region where the c.m.\ energy is much smaller than the photon
  virtuality can be described in a factorized form, as the convolution
  of a partonic handbag diagram and generalized distribution
  amplitudes which are new non-perturbative functions describing the
  exclusive fragmentation of a quark-antiquark pair into two
  hadrons. Scaling behavior and a selection rule on photon helicity
  are signatures of this mechanism. The case where $h$ is a pion is
  emphasized. }
\vskip 1.5\baselineskip
\end{center}

%%%%%%%%%%%%%%%%%%%%%%%%%%%%%%%%%%%%%%%%%%%%%%%%%%%%%%%%%%%%%%%%%%%%

\noindent
{\bf 1 Introduction.} Hadron production at small invariant mass or
even near threshold is usually quite inaccessible to a partonic
description. We advocate that a notable exception is provided by the
process $\gamma^\ast \gamma \to h \bar h $ with a highly virtual
photon, more precisely in the regime where the squared c.m.\ energy
$W^2$ is much smaller than the photon virtuality $Q^2$. We will here
focus on pairs of spin zero mesons, e.g.\ pions or kaons, but our
analysis can be extended to other final states such as $p \bar{p}$.

\begin{figure}
\begin{center}
  \leavevmode
  \epsfxsize 0.7\textwidth
  \epsfbox{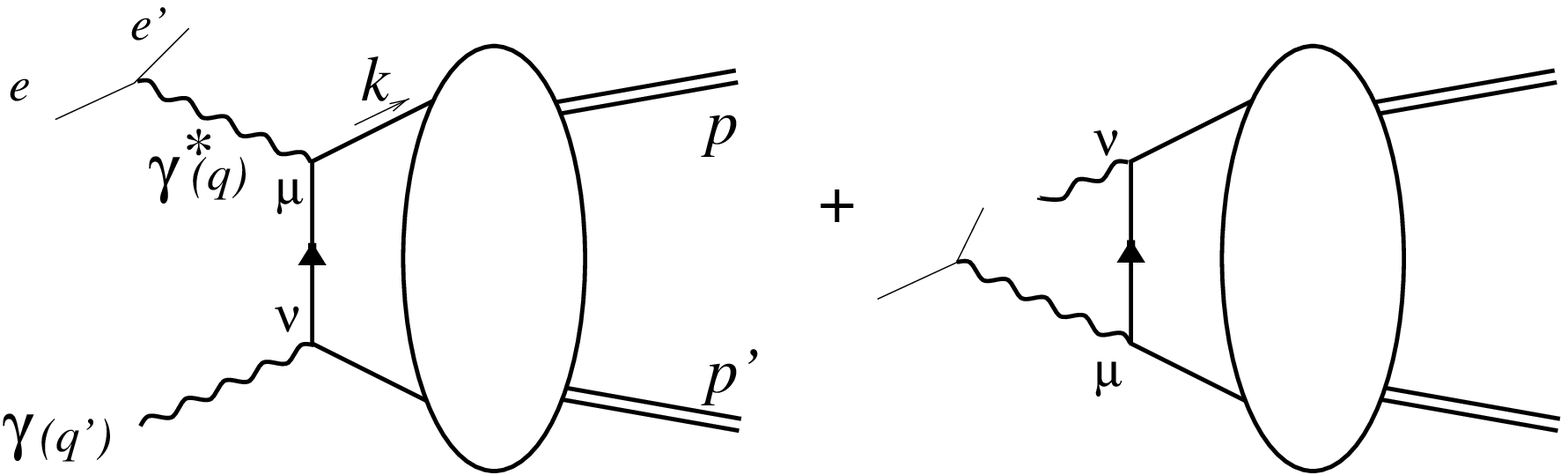}
\end{center}
\caption{\label{fig:process}Factorization of $\gamma^\ast \gamma \to h
  \bar{h}$ at tree level.}
\end{figure}

A natural point of reference for this study is the transition form
factor $\gamma^\ast \gamma \to \pi^0$ at large $Q^2$, which has been
extensively studied in QCD~\cite{BL,RR} and is considered to be one of
the best environments to study the pion light cone distribution
amplitude. We investigate a similar process, where a \emph{pair} of
neutral or charged pions is produced. The hard scattering,
$\gamma^\ast \gamma \to q \bar{q}$ at tree level, is essentially the
same, while the hadronic matrix element is a new non-perturbative
object describing the transition from partons to hadrons. This
mechanism is shown in Fig.~\ref{fig:process}.

Note that our reaction is related by crossing to deeply virtual
Compton scattering, $\gamma^\ast h \to \gamma h$, which at large $Q^2$
and small squared momentum transfer between the hadrons factorizes
into a hard photon-parton scattering and an off-diagonal parton
distribution~\cite{Ji,Rad}. The tree level diagrams are precisely the
crossed versions of those in Fig.~\ref{fig:process}. The $\gamma^\ast
\gamma$ and the Compton process share many features, in particular
their scaling behavior in $Q^2$ and a helicity selection rule for the
virtual photon.

\vskip\baselineskip
\noindent
{\bf 2 Kinematics.} In addition to the four-momenta shown in
Fig.~\ref{fig:process} we introduce $\PP = p+p'$ and $Q^2 = -q^2$,
$W^2 = \PP^2$, $t = (q-p)^2$, $\mm^2 = p^2$. Let us go to the
collision c.m.\ and define our $z$-axis along $\mathbf{q}$. In terms
of the angle $\theta_{\mathrm{cm}}$ between $\mathbf{q}$ and
$\mathbf{p}$ we may write
\begin{equation}
  \label{t}
  t = \mm^2 - {Q^2+W^2 \over 2} \left(1 - 
  \sqrt{1- {4 \mm^2 \over W^2}}\, 
  \cos\theta_{\mathrm{cm}}\right)  \eqpt
\end{equation}
Now we perform a boost along $z$ to the Breit frame and introduce
lightlike vectors $v = (1,0,0,1) /\sqrt{2}$ and $v' = (1,0,0,-1)
/\sqrt{2}$ which respectively set a ``plus'' and a ``minus''
direction. We then have
\begin{eqnarray}
q   &=& {Q\over\sqrt{2}}\,(v-v')  \eqcm \hspace{4em}
q'   =  {Q^2+W^2\over\sqrt{2}\,Q}\,v'  \eqcm \nonumber \\
\PP &=& {Q\over\sqrt{2}}\,v + {W^2\over\sqrt{2}\,Q}\,v'  \eqpt
\end{eqnarray}
We consider the kinematic region
\begin{equation}
  W^2 \ll Q^2  \eqcm
\end{equation}
where both $p$ and $p'$ have small minus and transverse components; to
describe their plus components we define the light cone fraction
\begin{equation}
\zeta = {p^+ \over P^+} = 1 - {\mm^2-t \over Q^2+W^2}  \eqcm
\end{equation}
which controls how they share the momentum $P$.  With~(\ref{t}) it is
easy to see that $\zeta$ and $1-\zeta$ cannot be smaller than $\mm^2 /
W^2$. This frame is suited to display the factorization of our process
into a hard, head-on collision of the two photons in which partons are
produced moving fast into the virtual photon direction, and a
long-distance process of hadronization. {}From usual power counting
arguments for Feynman graphs it is favorable to have the minimal
number of partons, that is two, between the hard scattering and the
soft hadron formation.

\vskip\baselineskip
\noindent
{\bf 3 Expression of the amplitude.} Let us now derive the tree level
expression of the $\gamma^\ast \gamma$ amplitude as a convolution of a
hard scattering amplitude $H$ and an amplitude $S$ for the soft
transition from partons to hadrons. Technically factorization means
that the behavior of $S$ enforces $k^2$ and $k\cdot \PP$ to be
negligible with respect to the hard scale $Q^2$. As a result the minus
and transverse components of $k$ can be neglected in $H$ and the
hadronic tensor can be written as an integral over the light cone
fraction $z = k^+ / \PP^+$ of the quark with respect to the hadronic
system:
\begin{eqnarray}
i T^{\mu\nu} &=& - \int\! d^4 x\, e^{-i q\cdot x} \,
    \out\langle h(p) \bar{h}(p') |\,
    T J_{\mathrm{em}}^\mu(x) J_{\mathrm{em}}^\nu(0) \,| 0 \rangle\in
    \nonumber \\ 
&=& \int\! d z\, H^{\mu\nu}_{\alpha\beta}(z,q,q') \,
    S_{\alpha\beta}^{\phantom{\mu}}(z,v',p,p') \eqcm
\end{eqnarray}
with the soft matrix element
\begin{eqnarray}
\label{soft-S}
S_{\alpha\beta} &=& {P^+ \over2\pi} \int\! dx^-\, 
  e^{-i z (P^+ x^-)} \,
%\cdot   \nonumber \\
%&& 
   \out\langle h(p) \bar{h}(p') |\, 
   \bar{\psi}_\alpha(x^- v')\psi_\beta(0) \,| 0 \rangle\in  \eqpt
\end{eqnarray}
Here we are working in the light cone gauge $A^+ = 0$, otherwise the
usual path ordered exponential appears between the quark fields. The
hard scattering is given by
\begin{eqnarray}
H_q^{\mu\nu} &=& {i e_q^2\over\sqrt{2}\, Q}\left\{
\left( g^{\mu \rho} v'^\nu + v'^\mu g^{\nu \rho}
       - g^{\mu\nu} v'^\rho \right) \gamma_\rho\,
     {2z-1\over z(1-z)} \right.  \nonumber \\
&& \left. \phantom{{i e_q^2\over\sqrt{2}\, Q}}
   - i \epsilon^{\mu\nu\rho\sigma} 
   \gamma_\rho \gamma_5
   \left(  v_\sigma - {v'_\sigma\over z(1-z)} \right)\right\}
\end{eqnarray}
for quark charge $e_q$, and selects the vector and axial vector part
of the soft amplitude. Note that the $v_\sigma$-term in the
antisymmetric part of $H^{\mu\nu}$ does not contribute at leading
power in $1/Q$. {}From now on we specialize to the case where $h$ is a
spin zero meson, then the axial part of $S$ vanishes due to parity
invariance and we are left with the vector part; note that this is
just opposite to the case of the $\gamma^\ast \gamma \to \pi^0$
transition form factor. Introducing for each quark flavor the
generalized distribution amplitude
\begin{equation}
S_{q, \alpha \beta}^{\phantom{\mu}}\,  \gamma^{+}_{\alpha \beta}
   =  \dist_q (z, \zeta, W^2) \, P^+
\end{equation}
we finally have
\begin{eqnarray}
\label{final-tensor}
T^{\mu\nu} &=& {1\over 2}
\left( v^\mu v'^\nu + v'^\mu v^\nu -g^{\mu\nu} \right) \, 
\sum_q e_q^2 \,
%\cdot  \nonumber \\
%&& 
   \int_0^1 dz\, {2z-1\over z(1-z)} \,
   \dist_q(z, \zeta, W^2)  \eqpt
\end{eqnarray}
Contracting with the photon polarization vectors we see that in order
to give a nonzero $\gamma^\ast \gamma \to h \bar{h}$ amplitude the
virtual photon must have the same helicity as the real one, in
particular it must be transverse. As in the case of virtual Compton
scattering this is a direct consequence of chiral invariance in the
collinear hard scattering process~\cite{DGPR}. We also find that the
$\gamma^\ast \gamma$ amplitude is independent of $Q^2$ at fixed
$\zeta$ and $W^2$, up to logarithmic scaling violation to be discussed
later.

There will of course be power corrections in $1/Q$ to this leading
mechanism. An example is the hadronic component of the real $\gamma$,
which might be modeled by vector meson dominance. {}From power
counting arguments one obtains that this contribution is suppressed,
as in virtual Compton scattering~\cite{Ji,Rad,Coll}. Our process can
also be treated within the operator product expansion~\cite{Wat},
which allows for a systematic analysis of higher twist effects.

\vskip\baselineskip
\noindent
{\bf 4 Generalized distribution amplitudes.} Although $S$ describes an
amplitude we have not time ordered the fields in its definition
(\ref{soft-S}): the time ordering can be left out because the
separation of the field operators is lightlike, as shown for ordinary
distribution amplitudes and off-diagonal or diagonal parton
distributions in~\cite{DG}. As a by-product of the proof one obtains
that our generalized distributions $\dist(z,\zeta,W^2$) are only
nonzero in the interval $0 < z < 1$, just as ordinary distribution
amplitudes.

Whereas time reversal invariance constrains ordinary distribution
amplitudes and parton distributions to be real valued functions up to
convention dependent phases, our generalized distributions are
\emph{complex}: time reversal interchanges incoming and outgoing
states, and $| h \bar{h} \rangle\out$ is different from $| h \bar{h}
\rangle\in$ since hadrons interact. We notice in (\ref{final-tensor})
that the hard scattering kernel at Born level is purely real so that
the imaginary part of the $\gamma^\ast \gamma$ amplitude is due to
$\mathrm{Im}\, \dist$; it corresponds to rescattering and resonance
formation in the soft transition from the partons to the final state
hadron pair.

Charge conjugation invariance provides a symmetry relation
\begin{equation}
\label{symmetry}
\dist(z, \zeta) = - \dist(1-z, 1-\zeta) \eqcm
\end{equation}
where for ease of writing we have dropped the argument $W^2$, so that
in the convolution (\ref{final-tensor}) one may replace
\begin{equation}
\dist(z, \zeta) \to {1 \over 2}
\Big[ \dist(z, \zeta) + \dist(z, 1 - \zeta) \Big]  \eqpt
\end{equation}
The amplitude is hence symmetric under $\zeta \leftrightarrow 1 -
\zeta$, i.e.\ under exchange of the hadron momenta, which reflects the
fact that the initial state $\gamma^\ast \gamma$ of the collision has
positive charge conjugation parity $C$. In the case where $h =
\bar{h}$, for instance if $h = \pi^0$, the corresponding
configurations are identical so that $\dist(z, \zeta)$ is even under
$\zeta \leftrightarrow 1 - \zeta$, and by virtue of (\ref{symmetry})
odd under $z \leftrightarrow 1 - z$.

As for parton distributions and ordinary distribution amplitudes there
are sum rules relating moments of $\dist$ to matrix elements of local
operators. Appropriately summed over quark flavors the moment $\int\!
d z\, \dist(z, \zeta)$ gives the timelike elastic form factor as
measured in $\gamma^\ast \to h \bar{h}$ and thus provides a constraint
on our new quantities. It is however inaccessible in our two-photon
process, which projects out the part of $\dist(z, \zeta)$ that is odd
under $z \leftrightarrow 1 - z$ as we see
in~(\ref{final-tensor}). This is clear because the elastic form factor
is $C$-odd while a $\gamma^\ast \gamma$ collision produces the
$C$-even projection of $h \bar{h}$. The moment $\int\! d z\, (2z-1)\,
\dist(z, \zeta)$ in contrast gives a matrix element of the quark part
of the energy-momentum tensor, which is $C$-even.

Let us point out some connections of our new distributions with other
quantities describing the partonic structure of hadrons. The physics
they involve goes beyond that of a $q\bar{q}$ distribution amplitude
of a meson: since two hadrons are formed the physics of $\dist$ does
not select their lowest Fock states; in this respect it is related to
ordinary parton distributions and to fragmentation functions. If on
the other hand $W$ is at or near the mass of a resonance with
appropriate quantum numbers, such as an $f_0$, it will contain physics
of the distribution amplitude for the resonance and of its decay into
$h \bar{h}$.

As already mentioned our distributions can also be viewed as the
crossed version of off-diagonal parton distributions~\cite{Ji,Rad},
and we remark that $\dist$ can be obtained from double
distributions~\cite{Rad} in the crossed channel. Notice also that
matrix elements like the ones defining $\dist$ have been considered
for multi-hadron production at large invariant
mass~\cite{BG-clusters}.

\vskip\baselineskip
\noindent {\bf 5 Evolution.}  QCD radiative corrections to the hard
scattering will as usual lead to logarithmic scaling violation. At
this point it is useful to remember the analogy of our process with
the $\gamma^\ast \gamma \to \pi^0$ transition form factor. The
evolution can be obtained from the hard scattering kernel alone and
remains the same if we replace $\dist$ with the distribution amplitude
of a meson with the appropriate quantum numbers, say an $f_0$. The
generalized distribution amplitudes thus follow the usual
Efremov-Radyushkin-Brodsky-Lepage (ERBL) evolution~\cite{ERBL} for a
meson,
\begin{equation}
\mu^2 {d \over d \mu^2}\, \dist(z, \zeta, W^2; \mu^2) =
\int_0^1\! dy\, V_+(z,y)\, \dist(y, \zeta, W^2; \mu^2)  \eqcm
\label{ERBL}
\end{equation}
where $\mu^2$ is the factorization scale and $V_+(z,y)$ the evolution
kernel.

The situation here is easier than in virtual Compton scattering, where
the evolution~\cite{Ji,Rad,VCS-evolution} is given by the
\emph{extended} ERBL kernel. In that case the hard subprocess involves
the parameter describing the kinematic asymmetry of the two hadrons,
whereas in our case the hard scattering and the evolution kernel are
independent of $\zeta$.

Up to now we have discussed generalized $q\bar{q}$ distribution
amplitudes. Beyond tree level in the hard scattering the hadron pair
can however also originate from two gluons. The generalized $q\bar{q}$
and $g g$ distributions mix under evolution and (\ref{ERBL}) becomes a
matrix equation. The corresponding mixing in the case of single
production of a pseudoscalar meson has been discussed
in~\cite{BG-evolution}. We remark that at one-loop level the
calculation of the $q\bar{q}$ diagonal element in the matrix evolution
kernel $V_+(z,y)$ is completely analogous to the standard one for
pseudoscalar and vector mesons and results in the same expression,
given in~\cite{ERBL}. As for the gluons, they are known to be
important in fragmentation and in parton distributions; following our
remark in the last section the generalized $g g$ distribution can
therefore be expected to be of the same order as the one for
$q\bar{q}$.

\vskip\baselineskip
\noindent
{\bf 6 Phenomenology.} $e^+ e^-$ high energy colliders such as $B\bar
B$ factories and LEP have a rich potential in $\gamma^\ast \gamma$
physics. Without elaborating on experimental issues we only remark
that the $\gamma^\ast \gamma$ c.m.\ is not the laboratory frame so
that our kinematics, even if $W$ is close to threshold, does not imply
slowly moving final state hadrons whose detection would be difficult.

In $e\gamma$ collisions the $\gamma^\ast \gamma$ process we want to
study competes with bremsstrahlung, where the hadron pair originates
from a virtual photon radiated off the lepton~\cite{Bud}.  This
process produces the pair in the $C$-odd channel and hence does not
contribute for $h = \bar{h}$, in particular not for $h = \pi^0$. Its
amplitude can be computed from the value of the timelike elastic form
factor measured in $e^+ e^- \to h \bar{h}$.

The interference between the $\gamma^\ast \gamma$ and bremsstrahlung
processes provides an opportunity to study the $\gamma^\ast \gamma$
contribution at \emph{amplitude} level. Thanks to the different charge
conjugation properties of the two processes their interference term
can be selected by the charge asymmetries
\begin{equation}
d\sigma( h(p) \bar{h}(p') ) - 
d\sigma( \bar{h}(p) h(p') )
\end{equation}
or
\begin{equation}
d\sigma (e^+ \gamma \to h \bar{h}) - 
d\sigma (e^- \gamma \to h \bar{h})  \eqcm
\end{equation}
while it drops out in the corresponding charge averages. We note that
bremsstrahlung has an amplitude with both real and imaginary parts,
especially at values of $W$ where it is dominated by vector meson
resonances, and in particular benefits from the $\rho$ up to $W$ of
order 1 GeV.

In a kinematical regime where bremsstrahlung is negligible a
comparison between charged and neutral pion pair yields will allow one
to study the breaking of isospin symmetry: applying this symmetry to
the transition from $q\bar{q}$ or $g g$ to a pair of pions in the
$C$-even channel one obtains
\begin{equation}
  T^{\mu\nu}( \pi^+(p) \pi^-(p') ) =
- T^{\mu\nu}( \pi^0(p) \pi^0(p') )
\end{equation}
for the hadronic tensor of the two-photon process.

We finally point out that using the methods of~\cite{DGPR} the angular
correlation between the leptonic and hadronic planes in the
$\gamma^\ast \gamma$ c.m.\ can be used to test the helicity selection
rule of the handbag mechanism and thus the dominance of leading twist
at finite values of $Q^2$.

\vskip\baselineskip
\noindent
{\bf 7 Conclusion.} We have exhibited a new instance of factorization
of long and short distance dynamics in a process accessible at
existing or planned $e^+ e^- $ or $e \gamma$ facilities. The
investigation of $\gamma^\ast \gamma \to h \bar{h}$ in the kinematical
domain $W^2 \ll Q^2$ provides a complement to the existing tools for
the study of hadrons in QCD.

The reaction studied and the generalized distribution amplitudes
introduced here are natural extensions of the $\gamma^\ast \gamma \to
\pi^0$ transition form factor and of the pion distribution amplitude,
which is of great importance in the QCD description of exclusive
reactions at large momentum transfer. Describing a $q \bar{q} \to h
\bar{h}$ or $g g \to h \bar{h}$ transition the new distribution
amplitudes contain more variables, specifying the invariant mass of
the two hadrons and their sharing of momentum. Note that they do not
select a lowest Fock state component of the hadrons.

The generalized distribution amplitudes may also be seen as quantities
obtained by crossing from the off-diagonal parton distributions,
discussed extensively in the framework of deeply virtual Compton
scattering and deep electroproduction of mesons.

Phenomenological aspects of the reaction presented here have been
sketched: charge asymmetries, interference with $h \bar{h}$-production
by bremsstrahlung, and angular correlations offer valuable help for
its experimental investigation. Once the validity of our leading-twist
analysis in a given region of $Q^2$ has been tested, the extraction of
the generalized distribution amplitudes for various mesons will be
possible; these new non-perturbative quantities should give us another
glimpse of how hadrons are made from partons.

\vskip\baselineskip
\noindent
{\bf Acknowledgments.} We wish to thank B. Meyer for a stimulating
discussion, and S.V. Mikhailov for helpful correspondence.


\begin{thebibliography}{99}
\vspace{8pt}
\setlength{\parskip}{6pt}

\bibitem{BL} G.P. Lepage and S.J. Brodsky, Phys.\ Rev.\ {\bf D22},
2157 (1980).

\bibitem{RR} S. Ong, Phys.\ Rev.\ {\bf D52}, 3111 (1995); R. Jakob,
P. Kroll and M. Raulfs, J. Phys. {\bf G22}, 45 (1996); P. Kroll and
M. Raulfs, Phys.\ Lett.\ {\bf B387}, 848 (1996); A.V. Radyushkin and
R.T. Ruskov, Nucl.\ Phys.\ {\bf B481}, 625 (1996); I.V. Musatov and
A.V. Radyushkin, Phys.\ Rev.\ {\bf D56}, 2713 (1997).

\bibitem{Ji} X. Ji, Phys.\ Rev.\ Lett.\ {\bf 78}, 610 (1997); Phys.\
  Rev.\ {\bf D55}, 7114 (1997).

\bibitem{Rad} A.V. Radyushkin, Phys.\ Lett.\ {\bf B380}, 417 (1996);
Phys. Rev. {\bf D56}, 5524 (1997).

\bibitem{DGPR} M. Diehl {\it et al.},  Phys.\ Lett.\ {\bf B411}, 193
(1997).

\bibitem{Coll} J.C. Collins, L. Frankfurt and M. Strikman, Phys.\
Rev.\ {\bf D56}, 2982 (1997); J.C. Collins and A. Freund,
hep-ph/9801262.

\bibitem{Wat} K. Watanabe, Progr.\ Theor.\ Phys.\ {\bf 67}, 1834
(1982).

\bibitem{DG} M. Diehl and T. Gousset, hep-ph/9801233, to appear in
Phys.\ Lett.\ {\bf B}.

\bibitem{BG-clusters} V.N. Baier and A.G. Grozin, Sov.\ J.\ Nucl.\
Phys.\ {\bf 35}, 899 (1982).

\bibitem{ERBL} G.P. Lepage and S.J. Brodsky, Phys.\ Lett.\ {\bf B87},
359 (1979); A.V. Efremov and A.V. Radyushkin, Phys.\ Lett.\ {\bf B94},
245 (1980).

\bibitem{VCS-evolution} J. Bl\"umlein, B. Geyer and D. Robaschik,
Phys.\ Lett.\ {\bf B406}, 161 (1997); A.V. Belitsky and D. M\"uller,
Phys.Lett. {\bf B417}, 129 (1998); A.V. Belitsky and A. Sch\"afer,
hep-ph/9801252.

\bibitem{BG-evolution} V.N. Baier and A.G. Grozin, Nucl.\ Phys.\ {\bf
  B192}, 476 (1981).

\bibitem{Bud} V.M. Budnev {\it et al.}, Phys.\ Rept.\ {\bf C15}, 181
(1975).

\end{thebibliography}
\end{document}